\PassOptionsToPackage{table}{xcolor}
\documentclass[sigconf]{acmart} %

\author{Denini Silva}
\affiliation{%
  \institution{Federal University of Pernambuco}
  \city{Recife}
  \state{PE}
  \country{Brazil}
}
\email{dgs@cin.ufpe.br}

\author{MohamadAli Farahat}
\affiliation{%
  \institution{North Carolina State University}
  \city{Raleigh}
  \state{NC}
  \country{USA}
}
\email{mfaraha3@ncsu.edu}

\author{Marcelo d'Amorim}
\affiliation{%
  \institution{North Carolina State University}
  \city{Raleigh}
  \state{NC}
  \country{USA}
}
\email{mdamori@ncsu.edu}

\renewcommand{\shortauthors}{Silva \etal}

\usepackage{comment}

\usepackage{lipsum}

\usepackage{booktabs}     %

\usepackage{graphicx}
\usepackage{subcaption}
\usepackage{placeins} %
\usepackage{float}     %
\usepackage{tcolorbox}
\tcbuselibrary{skins,breakable,listings}
\usepackage{listings}
\usepackage{xcolor}  %
\usepackage{array}
\lstdefinestyle{code}{
  basicstyle=\ttfamily\small,
  numbers=left, numberstyle=\scriptsize, numbersep=6pt,
  breaklines=true, breakatwhitespace=true,
  frame=single, framerule=0.4pt,
  tabsize=2, showstringspaces=false
}
\lstdefinestyle{cpp}{style=code, language=C++}
\lstdefinestyle{json}{style=code, language=}
\lstdefinestyle{txt}{style=code, language=}

\newtcblisting{codecard}[2][]{%
  enhanced, breakable,
  colback=white, colframe=black!12,
  boxrule=0.6pt, arc=2mm, left=1mm, right=1mm, top=1mm, bottom=1mm,
  title={#2}, fonttitle=\bfseries\footnotesize,
  listing only, listing options={#1}
}

\usepackage{tabularx}

\usepackage{wrapfig}
\usepackage{multirow}
\usepackage{multicol}
\usepackage{enumitem}
\usepackage{listings}
\usepackage{xspace}
\usepackage{subcaption}
\usepackage{pgf-pie}
\usepackage{pifont}

\usepackage{tikz}
\tikzset{>=latex}
\newcommand*\circled[1]{\tikz[baseline=(char.base)]{
            \node[shape=circle,draw,fill=black,text=white,inner
            sep=0.4pt] (char) {\small #1};}}

\newcommand{\cmark}{\color{blue}{\ding{51}}}
\newcommand{\xmark}{\color{red}{\ding{55}}}

\usepackage[framemethod=tikz]{mdframed}
\mdfdefinestyle{mpdframe}{
   frametitlebackgroundcolor   =black!15,
    frametitlerule              =true,
    roundcorner                 =3pt,
    middlelinewidth             =1pt,
    skipabove                   =\topskip,
    skipbelow                   =\topskip,
    innermargin                 =0.1cm, %
    outermargin                 =0.1cm,
    innerleftmargin             =0.1cm,
    innerrightmargin            =0cm,
    innertopmargin              =0.1cm,
    innerbottommargin           =0.1cm,
    align=center   
}

\newcommand{\DefMacro}[2]{%
   \expandafter\newcommand\csname rmk-#1\endcsname{#2}%
}
\newcommand{\UseMacro}[1]{\csname rmk-#1\endcsname}

\newcommand{\etal}{et al.}
\newcommand{\eg}{e.g.}
\newcommand{\ie}{i.e.}

\newcommand{\Comment}[1]{}

\newcommand{\Space}[1]{}

\newcommand{\MyPara}[1]{\vspace{1pt}\noindent\textbf{#1}.}

\newcommand{\Code}[1]{{\small\ifmmode{\texttt{#1}}\else$\texttt{#1}$\fi}}
\newcommand{\CodeIn}[1]{{\small\ifmmode{\mathtt{#1}}\else$\mathtt{#1}$\fi}}
\newcommand{\ColorBack}[1]{%
  \begingroup \setlength{\fboxsep}{0pt}%
}

\definecolor{gray}{RGB}{211,211,211}

\newcommand{\PaperTitle}{An Empirical Analysis of Cross-OS Portability\\ Issues in Python Projects}

\newcommand{\github}{GitHub\xspace}

\newcommand{\Contrib}[1]{$\star$#1}

\newcommand{\EQ}[1]{RQ#1\xspace}

\newcommand{\RQOne}{\EQ{1}:~How prevalent are portability problems in Python projects?}

\newcommand{\RQTwo}{\EQ{2}:~What are the causes, symptoms, and fixes of portability problems?}
\newcommand{\RQTwoOne}{\EQ{2.1}: What are the root causes of portability problems in Python?}
\newcommand{\RQTwoTwo}{\EQ{2.2}: What are the observable symptoms of portability issues?}
\newcommand{\RQTwoThree}{\EQ{2.3}: What are fix patterns for addressing portability issues?}

\newcommand{\RQThree}{\EQ{3}: How good are existing tools at detecting
and fixing portability issues?}

\newcommand{\RQFour}{\EQ{4}: How Do Developers Respond to Reported Portability Issues?}

\newcommand{\ourrepo}{\url{https://github.com/ncsu-swat/portability-issues-py}}

\newcommand{\NumProjectsTotal}{2{,}042}              %
\newcommand{\NumProjectsLearning}{900}                %
\newcommand{\NumProjectsReExecution}{500}             %
\newcommand{\NumProjectsIssueMining}{400}             %
\newcommand{\NumProjectsApplication}{1{,}142}         %

\newcommand{\NumProjectsWithOSDiffInTests}{56}        %
\newcommand{\NumTestsExecuted}{440,728}               %
\newcommand{\NumTestRunsWithOSDiff}{9,508}            %
\newcommand{\PercentDiscrepantTests}{2.16\%}          %
\newcommand{\PercentProjectsWithPortabilityReExecution}{11.2\%}  %

\newcommand{\NumIssuesMined}{2,617}                   %
\newcommand{\NumIssuesAnalyzed}{240}                  %
\newcommand{\NumIssuesRelated}{102}                   %
\newcommand{\NumProjectsWithOSDiffInIssues}{95}       %

\newcommand{\TotalProjectsWithPortabilityIssues}{151} %

\newcommand{\NumRootCauseCategories}{7}               %
\newcommand{\NumRootCauseSubcategories}{24}           %
\newcommand{\NumSymptomSignatures}{15}                %
\newcommand{\NumContextSymptoms}{9}                   %
\newcommand{\NumFixPatterns}{4}                       %

\newcommand{\NumCodeUsedLLM}{90}                      %
\newcommand{\NumCodeNonPortableByLLM}{30}             %
\newcommand{\NumCodePortableIssue}{30}                %
\newcommand{\NumCodePortableManual}{30}               %

\newcommand{\TotalPROpened}{33}
\newcommand{\TotalPRAccepted}{17}
\newcommand{\TotalPRRejected}{0}
\newcommand{\TotalPRInTestsCode}{17}
\newcommand{\TotalPRInProgramCode}{10}
\newcommand{\TotalPRInBothCode}{6}
\newcommand{\TotalPRAprovedPercent}{51.5\%}
\newcommand{\TotalPRInProgramPercent}{30.3\%}
\newcommand{\TotalPRInTestsPercent}{51.5\%}

\newcommand{\todo}[2][normal]{%
  \ifdefined\showtodos
    \ifstrequal{#1}{high}{\sethlcolor{red}}{}%
    \ifstrequal{#1}{normal}{\sethlcolor{yellow}}{}%
    \ifstrequal{#1}{low}{\sethlcolor{green}}{}%
    {\hl{\textbf{TODO:} #2}}%
    \marginpar{\textcolor{red}{\textbf{TODO}}}%
  \fi
}

\newif\ifshowtodos
\showtodostrue  %

\DefMacro{eq-single-performance}{\EQ{1}}
\DefMacro{eq-single-safety}{\EQ{2}}
\DefMacro{eq-evolution-stability}{\EQ{3}}
\DefMacro{eq-evolution-performance}{\EQ{4}}
\DefMacro{eq-evolution-safety}{\EQ{5}}
\DefMacro{Category}{Category}
\DefMacro{PolicyType}{Policy Type}
\DefMacro{Description}{Description}

\newenvironment{packed_itemize}{
\vspace{-0.1ex}
\begin{list}{\labelitemi}{\leftmargin=1.7em}
 \setlength{\itemsep}{0pt} \setlength{\parskip}{0pt} \setlength{\parsep}{0pt} \setlength{\headsep}{0pt} \setlength{\topskip}{0pt} \setlength{\topmargin}{0pt} \setlength{\topsep}{0pt} \setlength{\partopsep}{0pt}
 }{\end{list}
\vspace{-0.1ex}
}

\newenvironment{packed_enumerate}{
\vspace{-0.7ex}
\begin{enumerate}[label=\arabic*.,leftmargin=1.7em]%
 \setlength{\itemsep}{0.7pt}
 \setlength{\parskip}{0pt}
 \setlength{\parsep}{0pt}
 \setlength{\headsep}{0pt}
 \setlength{\topskip}{0pt}
 \setlength{\topmargin}{0pt}
 \setlength{\topsep}{0pt}
 \setlength{\partopsep}{0pt}
}{\end{enumerate}
\vspace{-0.7ex}
}

\definecolor{SubtleColor}{rgb}{0,0,.50}

\definecolor{DefensiveColor}{RGB}{65,105,225}      %
\definecolor{PortableColor}{RGB}{34,139,34}        %
\definecolor{NormColor}{RGB}{205,92,92}            %
\definecolor{EnvColor}{RGB}{138,43,226}            %

\definecolor{UniqueSymptomColor}{RGB}{200,230,255}     %
\definecolor{ContextSymptomColor}{RGB}{255,250,205}    %

\newcommand{\DefensivePattern}{\textcolor{DefensiveColor}{\textbf{Defensive checks}}}
\newcommand{\PortablePattern}{\textcolor{PortableColor}{\textbf{Portable APIs}}}
\newcommand{\NormPattern}{\textcolor{NormColor}{\textbf{Normalization}}}
\newcommand{\EnvPattern}{\textcolor{EnvColor}{\textbf{Environment handling}}}

\newcommand{\UniqueSymptom}[1]{\cellcolor{UniqueSymptomColor}#1}
\newcommand{\ContextSymptom}[1]{\cellcolor{ContextSymptomColor}#1}

\copyrightyear{2026}
\acmYear{2026}
\setcopyright{cc}
\setcctype{by}
\acmConference[MSR '26]{23rd International Conference on Mining Software Repositories}{April 13--14, 2026}{Rio de Janeiro, Brazil}
\acmBooktitle{23rd International Conference on Mining Software Repositories (MSR '26), April 13--14, 2026, Rio de Janeiro, Brazil}
\acmPrice{}
\acmDOI{10.1145/3793302.3793373}
\acmISBN{979-8-4007-2474-9/2026/04}

\begin{document}

\title{\PaperTitle{}}

\begin{abstract}
  \sloppy
  While Python is designed as a cross-platform language,
real-world applications encounter portability failures
when deployed across different operating systems.

We present the first large-scale empirical study of
cross-OS portability issues in Python, analyzing \NumProjectsTotal{}
open-source repositories using two complementary approaches:
systematic cross-OS test re-execution and manual analysis
of GitHub issues. Our cross-platform testing of
\NumProjectsReExecution{} projects reveals that
\PercentProjectsWithPortabilityReExecution{} exhibit
OS-dependent test failures.
Through systematic analysis of \NumIssuesAnalyzed{} GitHub issues,
we confirm \NumIssuesRelated{} genuine portability problems
spanning \NumProjectsWithOSDiffInIssues{} additional projects.
We develop a comprehensive taxonomy identifying
\NumRootCauseCategories{} primary failure categories—with
file/directory operations, process management, and library dependencies
being most prevalent—along with \NumRootCauseSubcategories{} 
distinct sub-categories, \NumSymptomSignatures{} diagnostic signatures, 
and \NumFixPatterns{} systematic repair patterns.
Our evaluation reveals
that existing static analysis tools provide minimal support for
portability detection, while large language models achieve 40--79\% 
accuracy in identifying issues and 50--77\% success in generating fixes
when provided with structured guidance. Through \TotalPROpened{} contributed pull requests,
we demonstrate practical applicability and developer acceptance 
(\TotalPRAccepted{} merged, zero rejected) of our findings.

This work establishes
the first comprehensive baseline for understanding and addressing
cross-OS portability issues in Python, providing actionable insights
for developers, tool designers, and the broader research community. 
\end{abstract}

\ccsdesc[500]{Software and its engineering~Software verification and validation}
\ccsdesc[500]{Software and its engineering~Functionality}

\keywords{Cross-Platform Portability Issues, Python Tests, Large Language Models.}

\maketitle

\section{Introduction}
\label{sec:intro}
\label{sec:sources}

A \emph{portability issue} refers to an observable difference in a
program’s behavior that arises from variations in the underlying
platform—for example, inconsistencies in API implementations across
platforms.
Python is widely regarded as a cross-platform language, powering
applications across diverse domains such as data science, web
development, and automation. It consistently ranks among the most
popular programming languages
today~\cite{stackoverflow2025,githuboctoverse2024,pypl2025}. However,
platform portability is not guaranteed. Mince et al.~\cite{mince2023},
for instance, reported that roughly 40\% of the functionality in
popular machine learning libraries (\eg{},
PyTorch~\cite{paszke2019pytorch}) behaves inconsistently across
hardware platforms such as CPUs and TPUs. Similar challenges occur at
the operating-system level; for example, the function
\CodeIn{os.geteuid()} is unavailable in Windows distributions of
Python~\cite{pythongeteuid}.

Portability issues in software projects often arise from dependencies
on platform-specific APIs, system libraries, and environment
assumptions. These issues commonly lead to deployment failures when
software is executed in environments differing from those used during
development or testing. Consequently, developers must spend
substantial time diagnosing and adapting code, which negatively
impacts productivity and delays releases. Prior work has shown that
the availability and usage of platform-specific APIs can directly
affect the portability and reliability of software
systems~\cite{job2024}. Industry reports further emphasize that
ensuring code portability requires deliberate cross-platform design,
testing, and maintenance
practices~\cite{kiuwancodeportability}. Systematic investigations
indicate that defining and quantifying portability remains a
persistent challenge in software engineering, with no widely adopted
metrics or automated measurement
frameworks~\cite{ghandorh2020systematic}. Moreover, empirical studies
have connected portability-related inconsistencies to unstable or
flaky test behavior, complicating continuous integration pipelines and
further burdening
developers~\cite{vahabzadeh2015empirical,flaky_tests_empirical}.
Container technologies such as Docker~\cite{docker} and Singularity
(now Apptainer)~\cite{apptainer} can address software portability
challenges by encapsulating code, dependencies, and execution
environments into reproducible units. They minimize configuration
discrepancies across platforms and facilitate consistent testing and
deployment. Nevertheless, despite the clear benefits of
containerization, most open-source Python projects remain
non-containerized. In our analysis of \NumProjectsTotal{}
\github\ repositories, we find that only 19.9\% included a Dockerfile
or Docker Compose configuration.

\sloppy
This paper presents a study of cross-OS portability issues in Python
projects. Specifically, we analyze \NumProjectsTotal{} repositories to assess the prevalence of such issues,
their characteristics (e.g., symptoms, root causes, and fixes), the
effectiveness of existing tools (e.g., static analyzers and large
language models) in detecting and repairing them, and developers'
reactions to the \TotalPROpened{} pull requests we submitted to address portability
issues in their code. Our analysis reveals that these
portability failures, while widespread (\PercentProjectsWithPortabilityReExecution{} of tested projects), are not chaotic or
unpredictable. Instead, they cluster around \NumRootCauseCategories{} main root-cause categories and \NumRootCauseSubcategories{} 
well-defined sub-categories—with file and directory operations, process management, and library dependencies
being most prevalent. Results indicate
that a combination of test re-execution and LLMs can complementarily 
detect issues, with LLMs achieving 40--79\% accuracy in detection and 50--77\% success in repair when provided with structured guidance.
We conclude with a set of lessons we learned.

This paper makes the following contributions:
\begin{packed_itemize}
  \item[\Contrib] A study about the prevalence of portability issues that use
    two complementary methods to identify issues: test re-execution
    and mining of issues with subsequent manual analysis;
  \item[\Contrib] A detailed categorization of symptoms, root causes, and fix
    patterns of portability issues;
  \item[\Contrib] An evaluation of the ability of static analyses and LLMs to detect and repair portability issues;
  \item[\Contrib] An evaluation about the reaction of developers about the pull
    requests we open to fix portability issues in their
    \github\ projects;
  \item[\Contrib] A dataset of portability issues across \NumProjectsTotal{}
    Python projects, including test re-execution results, GitHub issues, 
    code examples, and LLM evaluation data.
\end{packed_itemize}

Our artifacts, including code and data, are publicly available at \textbf{\ourrepo{}}.

\section{Projects and Questions}
\label{sec:methodology}

This section describes the projects we used (\ref{sec:projects}) the
research questions we posed (\ref{sec:questions}).

\subsection{Projects}
\label{sec:projects}

\noindent
\textbf{Dataset.} We identified candidate repositories through a broad 
\github{} API query, then filtered this pool by verifying Python was the 
predominant language and that repositories contained executable tests with 
at least one passing test. From this set, we sampled \NumProjectsTotal{} 
projects, a size both substantial and feasible for cross-platform 
execution, and randomly partitioned them for our analyses.

\textbf{Dataset partitioning.} We partitioned the dataset to balance human 
effort with obtaining a representative sample. We allocated 
\NumProjectsLearning{} projects to a \emph{learning set} for identifying 
and categorizing portability issues through detailed analysis. This set 
was divided by detection method: \NumProjectsReExecution{} projects 
(nearly 25\% of the total) through cross-OS test re-execution, and 
\NumProjectsIssueMining{} projects (an amount realistically inspectable 
manually) through systematic \github{} issue mining. These subsets are 
non-overlapping, ensuring independent observations. We extract our
taxonomy from these \NumProjectsLearning{} projects (=\NumProjectsReExecution{}+\NumProjectsIssueMining{}). The remaining 
\NumProjectsApplication{} projects (=\NumProjectsTotal{}-\NumProjectsApplication{})  constitute an \emph{application set} 
for validation through pull requests.

\begin{table}[H]

\centering
\caption{Stats on \NumProjectsTotal{} evaluated projects: no. of tests (\#Tests), size (SLOC), no. of commits (\#Sha), age in years (Age), no. of \github{} stars ($\star$).\label{tab:project-stats}}
\begin{tabular}{lrrrrr}
\hline
Statistic & \#Tests & SLOC & \#Sha & Age & $\star$ \\
\hline
Mean & 572.51 & 41,635.39 & 2,883.25 & 6.24 & 6,147.88 \\
Median & 72 & 7,478 & 426.50 & 5.87 & 83.50 \\
Min & 1 & 2 & 2 & 0.3 & 2 \\
Max & 36,588 & 3,237,895 & 194,153 & 17.27 & 366,472 \\
Sum & 1,169,056 & 85,019,457 & - & - & - \\
\hline
\end{tabular}
\end{table}

Table~\ref{tab:project-stats} summarizes key characteristics of the
\NumProjectsTotal{} evaluated projects.  On average, each project comprises
approximately 573 tests and 41.6K lines of Python code (SLOC), with a
median of 72 tests and 7.5K SLOC, indicating a distribution skewed by
a few very large systems. Regarding repository activity, projects have
on average 2.9K commits, a median of 427, and span a mean lifetime of
6.2 years. The number of \github{} stars varies widely, with an
average of 6.1K, a median of only 84, and a maximum of $\sim$366K,
reflecting the presence of both niche and highly popular projects.

Overall, the dataset covers a diverse range of software systems, from
small-scale repositories with only a handful of commits to large,
long-lived, and widely adopted projects.

\subsection{Research Questions and Methodology}
\label{sec:questions}

We aim to answer the following key research questions:

\sloppy \MyPara{\RQOne{}}~This question aims to assess how frequently
portability problems occur in practice. If such issues are very rare,
or if developers generally disregard them, the study’s practical value
would be limited. To address this question, we analyze issue reports
and execute tests in the wild across multiple virtual machines.

\vspace{0.5ex}
\MyPara{\RQTwo{}}~Assuming that identifying and
resolving portability issues is an important problem, it is also
essential to understand how these issues manifest (i.e., their
symptoms), what their root causes are, and how they are typically
repaired. For instance, automated solutions become more feasible when
a small number of recurring fix patterns can be observed and documented.

\vspace{0.5ex}
\MyPara{\RQThree{}}~This question examines the extent to which
existing tools (e.g., static analyzers and LLMs) can detect and fix
portability problems. Re-executing test suites across different OSes
(e.g., through virtual machines) can reliably identify portability
issues, but it is limited by the ability of the existing tests to
exercise the affected code locations.

\vspace{0.5ex}
\MyPara{\RQFour{}}~This question aims to further understand technical
and social dimensions of portability issues, namely, what are the
common fixes developers employ and how developers react to pull
requests we submit to fix portability issues in their projects.

\section{Results}
\label{sec:results}

This section provides answers to the posed research questions.

\subsection{\textbf{Answering} \RQOne{}}
\label{rq1}

\begin{table}[t!]
  \centering
  \caption{\label{tab:dataset}Prevalence of portability problems.}
  \vspace{-1ex}
  \begin{subtable}[t]{\linewidth}
    \centering
    \caption{\label{tab:dataset-a}Problems detected with test re-execution.}
    \begin{tabular}{lr}
    \toprule
     Number of projects selected & \NumProjectsReExecution{} \\
    \hspace{2ex} Number of projects with OS differences & \NumProjectsWithOSDiffInTests{} \\    
     Number of tests executed & \NumTestsExecuted{} \\
    \hspace{2ex} Number of test runs with OS differences & \NumTestRunsWithOSDiff{} \\
    \bottomrule
    \end{tabular}
  \vspace{1ex} %
  \end{subtable}
  \begin{subtable}[t]{\linewidth}
    \centering
    \caption{\label{tab:dataset-b}Problems detected in issues with manual inspection.}
    \begin{tabular}{lr}
      \toprule
       Number of projects selected & \NumProjectsIssueMining{} \\
      \hspace{2ex} Number of projects with OS differences & \NumProjectsWithOSDiffInIssues{} \\      
       Number of issues mined & \NumIssuesMined{} \\
      \hspace{2ex} Number of issues analyzed & \NumIssuesAnalyzed{} \\
      \hspace{4ex} Number of issues documenting OS differences & \NumIssuesRelated{}\\
      \bottomrule
    \end{tabular}
  \end{subtable}
\end{table}

\subsubsection{Methods}
\label{sec:methods}

We use two complementary methods to identify portability issues. The 
first method, \emph{cross-OS test re-execution}, systematically reveals behavioral 
differences across platforms but it is inherently incomplete as tests may 
not cover important code paths. The second method, \emph{issue mining}, 
complements test re-execution by including developer perspectives and 
discussions, though it is costly as it requires manual analysis of 
discussion threads to rule out spurious cases. We discuss both methods 
in detail below.

\textbf{Cross-OS test re-execution.} For each project we run the test suite
on multiple operating systems and record divergent outcomes. We then
triage failures via logs and small code inspections, mapping each
instance to a concrete category (i.e., a classification of the root cause 
of the portability issue, such as file handling, missing APIs, or library 
dependencies; the full taxonomy is presented in answering \RQTwo{}). We 
execute tests on ephemeral virtual machines provided by GitHub-hosted
runners~\cite{github_actions_runners}. Each virtual machine is
automatically provisioned with 4~CPUs and 16~GB of RAM. We implemented
a single GitHub Actions workflow (YAML) that uses a matrix strategy to
run the test-suite on Ubuntu 24.04 LTS, macOS 15, and Windows Server
2025 (each job runs in a fresh VM image). Each job installs
dependencies, runs the Python tests with
\texttt{pytest}~\cite{pytest_tool} and the \texttt{pytest-json-report}
plugin~\cite{pytest_json_report} to create JSON reports, and finally
uploads the JSON files as workflow artifacts for later analysis. This
approach ensures reproducible, parallel cross-platform testing while
GitHub manages VM provisioning and maintenance.

\textbf{Issues Mining.} To complement test-based findings, we selected
\NumProjectsIssueMining{} projects from the learning set and mined
\github{} issues and pull requests from them. We wrote a ``candidate finder''
combining multiple \emph{types} of keyword:  
(a) OS/platform indicators (e.g., Windows, Linux, macOS, specific distros/architectures),  
(b) failure/fix language (e.g., fails, error, bug, fix, workaround),  
(c) testing/CI context (pytest, CI, \github{} Actions), and  
(d) common portability causes (e.g., path separators, chmod/permissions, encodings/UTF-8, dynamic libraries like \texttt{.dll}/\texttt{.so}).  
We combine these keywords with ``OR'' within each type and ``AND''
across types. We searched titles, bodies, and comments, with extra
weight to title and close-proximity matches. We employ a lightweight
triage (brief summaries and negative filters for off-topic mentions)
to filter spurious results. Candidate issues were normalized into
consistent records (project, link, date, summary, compact tags like
OS=, FIX=, TEST=, CAUSE=), duplicates removed, and full text (title,
description, comments) archived for analysis.

\subsubsection{Results}

We applied both methods described in Section~\ref{sec:methods} to the 
learning set. Table~\ref{tab:dataset} reports on the prevalence of 
portability problems in \github{} Python projects. 
Table~\ref{tab:dataset-a} details results from test re-execution, while
Table~\ref{tab:dataset-b} details results from issue mining. 

Table~\ref{tab:dataset-a} shows that of the \NumProjectsReExecution{} projects 
analyzed, we found discrepant behavior in \NumProjectsWithOSDiffInTests{} 
(\PercentProjectsWithPortabilityReExecution{}) of them, \ie{}, projects where 
test suites had at least one test manifesting OS differences. These projects 
included a total of \NumTestsExecuted{} test cases, of which \NumTestRunsWithOSDiff{} 
showed discrepant outcomes, corresponding to \PercentDiscrepantTests{}.

Table~\ref{tab:dataset-b} shows the results from issue mining. 
From the \NumProjectsIssueMining{} projects selected (see Section~\ref{sec:methods}), 
we mined a total of \NumIssuesMined{} candidate issues. From this set, we randomly 
sampled \NumIssuesAnalyzed{} of the most recent issues for manual inspection. 
Through careful manual analysis, we confirmed that \NumIssuesRelated{} of 
these were genuine portability issues, spanning \NumProjectsWithOSDiffInIssues{} 
distinct projects. Combined with the test re-execution results, we identified
\textbf{\TotalProjectsWithPortabilityIssues{}} Python projects with portability issues 
considering both methods.

\begin{tcolorbox}[boxrule=0pt,sharp corners,boxsep=2pt,left=2pt,right=2pt,top=2.5pt,bottom=2pt]
\textbf{RQ\textsubscript{1} (prevalence)}: We find that portability
issues are relatively prevalent. For example,
\PercentProjectsWithPortabilityReExecution{} of the
\NumProjectsReExecution{} projects we analyzed with cross-OS test
re-execution have portability problems.  Additionally, our issue
mining approach shows that \NumProjectsWithOSDiffInIssues{} of the
projects we analyzed have genuine portability issues.
\end{tcolorbox}

\subsection{\textbf{Answering} \RQTwo{}}

\subsubsection{Method}~We conducted a qualitative study to categorize
the root causes, symptoms, and the fixes of portability problems.

We used cross-OS test re-execution and the analysis of issues to find
root causes and symptoms. For fixes, we needed to rely on the
discussion in the issue and the corresponding fix commits. In cross-OS
test re-execution, we analyzed error messages, stack traces, and the
failing test code to understand the root causes and symptoms. For
issue mining, we analyzed issue titles, descriptions, developer
discussions, and comments to understand the root causes and symptoms.
Through iterative open coding~\cite{strauss1990basics} across both
data sources, we identified recurring patterns and grouped instances
with similar underlying causes. 
The coding process was conducted independently 
by two co-authors, with regular discussion to resolve 
ambiguities and refine category definitions. To ensure reliability, we computed 
inter-coder agreement on a subset of the data, achieving 96.4\% agreement with a 
Cohen's kappa of 0.89, reflecting almost perfect inter-coder reliability. During the later phases of the coding process,
new issues no longer produced new categories, indicating that the taxonomy 
had reached theoretical saturation. 

The resulting taxonomy of root causes consists of \textbf{\NumRootCauseCategories{}}
high-level categories (e.g., FILE, PROC, LIB) with \textbf{\NumRootCauseSubcategories{}}
distinct sub-categories. For each instance, we documented observable
symptoms (error messages, behaviors) and examined corresponding fixes
in merged pull requests or commits to identify common repair
patterns. We identified \textbf{\NumSymptomSignatures{}} unique symptom signatures and 
\textbf{\NumContextSymptoms{}} context-dependent symptoms across the \NumRootCauseSubcategories{} 
sub-categories, and \textbf{\NumFixPatterns{}} general fix patterns that collectively 
address all observed portability problems. Table~\ref{tab:data_sources} summarizes the 
primary data sources used to identify each component of our taxonomy. Root-cause
categories were derived from patterns observed across both test
failures and issue discussions.  Symptom signatures were primarily
extracted from test re-execution, where concrete error messages and
stack traces provide direct evidence of portability problems. Fix
patterns were identified by examining code changes and discussions in
both test-related commits and issue-related pull requests.

\begin{table}[h!]
\centering
\caption{Data sources for portability taxonomy components.}
\label{tab:data_sources}
\begin{tabular}{lcc}
\toprule
\textbf{Component} & \textbf{Test Re-execution} & \textbf{Issue Mining} \\
\midrule
Root-cause categories & \cmark & \cmark \\
Symptoms & \cmark & \xmark \\
Fix patterns & \cmark & \cmark \\
\bottomrule
\end{tabular}
\end{table}

In the following subsections, we elaborate on the characteristics of
 portability problems: root-causes, symptoms, and fixes.

\subsubsection{\textbf{Answering \RQTwoOne{}}}
\label{rq2-1}

~

\noindent Table~\ref{tab:categories} presents our taxonomy of portability issues 
organized by root-causes. Each row groups related sub-categories
under a main category. The columns show: (i) the root-cause category name and acronym,
(ii) specific sub-categories within that category,
(iii) a brief description of each sub-category,
(iv) the number of distinct projects where the issue appeared in test execution,
(v) the number of distinct projects where the issue appeared in mined \github{} issues, and 
(vi) the total count across both sources. 

\begin{table*}[t]
    \centering
    \caption{Taxonomy of portability issues showing root-cause categories, sub-categories, brief descriptions, and number of distinct projects observed in test execution and mined \github{} issues.}
    \label{tab:categories}
    \scriptsize
    \setlength{\tabcolsep}{3pt}
    \renewcommand{\arraystretch}{1.15}
    \begin{tabularx}{\textwidth}{
    >{\raggedright\arraybackslash}p{0.18\textwidth}
    >{\raggedright\arraybackslash}X
    >{\raggedright\arraybackslash}p{0.23\textwidth}
    >{\centering\arraybackslash}m{0.08\textwidth}
    >{\centering\arraybackslash}m{0.08\textwidth}
    >{\centering\arraybackslash}m{0.07\textwidth}
}
    \toprule
    \textbf{Root-cause category (Acronym)} & \textbf{Sub-categories} & \textbf{Description} & \textbf{\# of projects in tests } & \textbf{\# of projects in issues} & \textbf{Total} \\
    \midrule
    
    \multirow[t]{6}{*}{File and Directories (FILE)} 
    & Path separators & Forward vs. backslash differences & 7 & 27 & \multirow{6}{*}{62} \\
    & Line endings & CRLF vs. LF mismatch & 1  & 5 &                      \\
    & Case sensitivity & Filename case handling varies & 1  & 0 &                      \\
    & File locking & Windows locks open files & 7 & 4 &                      \\
    & Encoding & UTF-8 not default everywhere & 2 & 7 &                      \\
    & Filesystem block size & Block size assumptions differ & 0 & 1 &                      \\
    \midrule
    
    \multirow[t]{3}{*}{Process and Signals (PROC)}
    & Shell command execution & Shell behavior differs across OS & 5 & 14 & \multirow{3}{*}{25} \\
    & Missing signals & Signals unavailable on Windows & 3 & 2 &                      \\
    & Address already in use & Port binding conflicts vary & 1 & 0 &                      \\
    \midrule
    
    \multirow[t]{4}{*}{Library Dependencies (LIB)}
    & Platform-specific libraries (e.g., fcntl) & Unix-only libraries missing elsewhere & 2  & 1 & \multirow{4}{*}{24} \\
    & Missing libraries & Dependencies not universally available & 2 & 11 &                      \\
    & Dynamic library loading & DLL vs. SO file issues & 0 & 5 &                      \\
    & Binary wheel mismatch & Architecture-specific wheel incompatibilities & 0 & 3 &                      \\
    \midrule
    
    \multirow[t]{2}{*}{API Availability (API)}
    & Methods: os.uname(), os.geteuid(), os.getuid(), os.getpgid() & OS methods missing on Windows & 13 & 2 & \multirow{2}{*}{17} \\
    & Modules: readline, resource & Optional modules not everywhere & 0 & 2 &                      \\
    \midrule
    
    \multirow[t]{4}{*}{Environment and Display (ENV)}
    & No display in GitHub CI (linux) & Headless CI lacks display server & 3 & 2 & \multirow{4}{*}{14} \\
    & GUI differences & Window manager behavior varies & 0 & 4 &                      \\
    & Missing curses & Curses unavailable on Windows & 1 & 0 &                      \\
    & Terminal capabilities & Terminal features differ across systems & 2 & 2 &                      \\
    \midrule
    
    \multirow[t]{3}{*}{Permissions and Limits (PERM)}
    & Permission & Permission models differ by OS & 3 & 1 & \multirow{3}{*}{7} \\
    & File descriptor limits & Resource limits vary across systems & 0 & 1 &                     \\
    & Symlink privileges & Windows requires admin for symlinks & 1 & 1 &                     \\
    \midrule
    
    \multirow[t]{2}{*}{System Information (SYS)}
    & /proc filesystem & /proc missing on macOS/Windows & 1 & 0 & \multirow{2}{*}{2} \\
    & Timezone database & Timezone data not always installed & 1 & 0 &                      \\
    \midrule
    \textbf{Total} & - & - & \textbf{\NumProjectsWithOSDiffInTests{}} & \textbf{\NumProjectsWithOSDiffInIssues{}} & \textbf{\TotalProjectsWithPortabilityIssues{}} \\
    \bottomrule
    \end{tabularx}
\end{table*}

We identify seven main root-cause categories that explain why portability problems occur:

\begin{packed_enumerate}
\item \textbf{File and Directories (FILE):} The most prevalent category (62 projects) stems
 from OS differences in file operations. Key issues include path separator differences
  (backslash vs. forward slash), line ending mismatches (\CodeIn{CRLF} vs. \CodeIn{LF}),
   file locking semantics, and text encoding defaults.

\item \textbf{Process and Signals (PROC):} This category (25 projects) captures failures
 due to OS process management differences. For example, shell execution varies across platforms
  (\CodeIn{cmd.exe} vs. \CodeIn{bash}), Windows lacks many Unix signals
   (\CodeIn{SIGHUP}, \CodeIn{SIGKILL}), and port binding produces different error codes.

\item \textbf{Library Dependencies (LIB):} These failures (24 projects) occur when code
 assumes platform-specific libraries. Issues include Unix-only modules (e.g., \CodeIn{fcntl}),
  missing dependencies, different dynamic library extensions (\CodeIn{.dll} vs \CodeIn{.so} vs
   \CodeIn{.dylib}), and architecture-specific binary wheels.

\item \textbf{API Availability (API):} This category (17 projects) covers Python APIs
 not universally available. Many OS module methods are Unix-specific (\CodeIn{os.uname()},
  \CodeIn{os.geteuid()}, \CodeIn{os.getpgid()}) and fail on Windows, while optional modules
   like readline and resource may be missing.

\item \textbf{Environment and Display (ENV):} These problems (14 projects) 
arise from runtime environment differences. For example, headless CI lacks display servers 
for GUI testing, window managers behave differently, terminal capabilities vary, 
and the curses module is unavailable on Windows (not included in standard Python distributions).

\item \textbf{Permissions and Limits (PERM):} These issues (7 projects) reflect OS
 differences in access control. For example, permission models differ (Unix chmod versus Windows ACLs),
  file descriptor limits vary, and symbolic link creation requires admin privileges on Windows.

\item \textbf{System Information (SYS):} This category (2 projects) includes failures
 from platform-specific system information mechanisms. For example, the /proc filesystem exists only on
  Linux, and timezone databases may not be installed by default.
\end{packed_enumerate}

\begin{tcolorbox}[boxrule=0pt,sharp corners,boxsep=2pt,left=2pt,right=2pt,top=2.5pt,bottom=2pt]
\textbf{RQ\textsubscript{2.1} (root causes)}: We identified \NumRootCauseCategories{} root-cause categories
 with \NumRootCauseSubcategories{} sub-categories. File and Directories (FILE) is most prevalent
  (62 projects), followed by Process and Signals (25 projects) and Library Dependencies (24 projects).
\end{tcolorbox}

\subsubsection{\textbf{Answering \RQTwoTwo{}}}
\label{rq2-2}

\noindent Understanding how portability issues manifest during
execution is essential for rapid diagnosis and triaging. In this
section, we examine the observable symptoms of portability
problems—the concrete error messages, exceptions, and behavioral
differences that appear in test logs, and CI
output. Table~\ref{tab:symptoms} maps the root causes (of portability
issues) to the corresponding symptom signature and fix pattern.

\noindent\begin{table*}[t]
    \centering
    \caption{Symptom signatures and general fix patterns for portability sub-categories.
     Symptom cells are color-coded: \colorbox{UniqueSymptomColor}{unique signatures} enable immediate diagnosis;
     \colorbox{ContextSymptomColor}{context-dependent} require code inspection.
     Fix patterns are color-coded to show reuse across 
     categories: \EnvPattern{}, \DefensivePattern{}, \NormPattern{}, \PortablePattern{}.}
    \label{tab:symptoms}
    \scriptsize
    \setlength{\tabcolsep}{3pt}
    \renewcommand{\arraystretch}{1.15}
    \begin{tabularx}{\textwidth}{
        >{\raggedright\arraybackslash}m{0.06\textwidth}
        >{\raggedright\arraybackslash}m{0.13\textwidth}
        >{\raggedright\arraybackslash}X
        >{\raggedright\arraybackslash}m{0.31\textwidth}
    }
    \toprule
    \textbf{Root-cause category} & \textbf{Sub-categories} & \textbf{Symptom (verbatim error / message)} & \textbf{General Fix Pattern} \\
    \midrule
    \textbf{FILE} &
    Path separators & \ContextSymptom{Context-dependent: hardcoded paths fail silently or with generic file errors} & \NormPattern{} — normalize paths with pathlib \\
    & Line endings & \UniqueSymptom{\texttt{AssertionError: `\textbackslash r\textbackslash n' != `\textbackslash n'}} & \NormPattern{} — normalize line endings \\
    & Case sensitivity & \ContextSymptom{Context-dependent: file exists but name differs only in case} & \NormPattern{} — case-normalize filenames \\
    & File locking & \UniqueSymptom{\texttt{PermissionError: [WinError 32] The process cannot access the file because it is being used by another process} (Windows)} & \EnvPattern{} — use context managers, explicit close \\
    & Encoding & \UniqueSymptom{\texttt{UnicodeDecodeError: `charmap' codec can't decode byte 0x\{XX\} in position N}} & \NormPattern{} — set explicit UTF-8 encoding \\
    & Filesystem block size & \ContextSymptom{Context-dependent: tests assert exact byte counts that vary by OS} & \NormPattern{} — normalize computations avoid assumptions \\
    \midrule
    \textbf{PROC} &
    Shell command execution & \ContextSymptom{Context-dependent: command syntax or exit codes differ across shells} & \PortablePattern{} — use subprocess.run shell=False \\
    & Missing signals & \UniqueSymptom{\texttt{AttributeError: module `signal' has no attribute `SIGHUP'} (or SIGKILL, SIGTERM variants)} & \DefensivePattern{} — guard signals per platform \\
    & Address already in use & \UniqueSymptom{\texttt{OSError: [Errno 98] Address already in use} (Linux) / \texttt{[WinError 10048]} (Windows)} & \DefensivePattern{} + \EnvPattern{} — catch OS-specific errors, use port 0 \\
    \midrule
    \textbf{LIB} &
    Platform-specific libraries& \UniqueSymptom{\texttt{ModuleNotFoundError: No module named `fcntl'} (Windows)} & \DefensivePattern{} — conditional import provide fallback \\
    & Missing libraries & \UniqueSymptom{\texttt{ModuleNotFoundError: No module named `\{library\_name\}'}} & \DefensivePattern{} — optional import with fallback \\
    & Dynamic library loading & \UniqueSymptom{\texttt{OSError: [WinError 126] The specified module could not be found} (Windows)} & \DefensivePattern{} — guard loading prefer pure-Python \\
    & Binary wheel mismatch & \ContextSymptom{Context-dependent: installation succeeds but runtime fails with ABI errors} & \EnvPattern{} — install platform-appropriate wheel \\
    \midrule
    \textbf{API} &
    Methods (e.g., os.uname) & %
    \UniqueSymptom{\texttt{AttributeError: module `os' has no attribute `\{method\}'} (Windows)} & \PortablePattern{} + \DefensivePattern{} + \EnvPattern{} — use platform.uname, hasattr guards, OS detection \\
    & Modules (e.g., readline)  & %
    \UniqueSymptom{\texttt{ModuleNotFoundError: No module named `\{module\}'} (Windows)} & \DefensivePattern{} — guard import use alternatives \\
    \midrule
    \textbf{ENV} &
    No display in GitHub CI & \UniqueSymptom{\texttt{TclError: no
        display name and no \$DISPLAY environment variable} (Linux)} & \EnvPattern{} — skip GUI tests in CI \\
    & GUI differences & \ContextSymptom{Context-dependent: widgets render or behave differently} & \EnvPattern{} — skip or mock GUI features \\
    & Missing curses & \UniqueSymptom{\texttt{ModuleNotFoundError: No module named `\_curses'} (Windows)} & \DefensivePattern{} — conditional import with stub fallback \\
    & Terminal capabilities & \ContextSymptom{Context-dependent: color codes appear as text or formatting ignored} & \EnvPattern{} — detect capabilities degrade gracefully \\
    \midrule
    \textbf{PERM} &
    Permission & \ContextSymptom{Context-dependent: depends on operation (chmod, file access, etc.)} & \DefensivePattern{} + \EnvPattern{} — check permissions, skip restricted tests \\
    & File descriptor limits & \UniqueSymptom{\texttt{OSError: [Errno 24] Too many open files}} & \EnvPattern{} — close files promptly, raise limits in CI \\
    & Symlink privileges & \UniqueSymptom{\texttt{OSError: [WinError 1314] A required privilege is not held by the client} (Windows)} & \EnvPattern{} — detect support and use copies \\
    \midrule
    \textbf{SYS} &
    /proc filesystem & \ContextSymptom{Context-dependent: \texttt{/proc}    paths fail on macOS/Windows with generic FileNotFoundError} & \PortablePattern{} — use psutil or platform APIs \\
    & Timezone database & \UniqueSymptom{\texttt{zoneinfo.ZoneInfoNotFoundError: `No time zone found with key \{...\}'}} & \NormPattern{} — install tzdata or fallback \\
    \bottomrule
    \end{tabularx}
    \end{table*}   

\MyPara{Symptom categories}~We categorize symptoms into two types based on 
their diagnostic value (color-coded in Table~\ref{tab:symptoms}): 
\textbf{(i) Unique signatures} are error messages 
that unambiguously identify a specific portability sub-category, enabling 
immediate diagnosis from logs alone; and \textbf{(ii) Context-dependent symptoms} 
are manifestations that require additional code inspection or contextual understanding 
to distinguish from other failure modes.

\MyPara{Unique signatures}~The majority of sub-categories (15 out of 24, 62.5\%) 
produce unique, actionable error signatures. These include missing API methods 
(\eg{}, \texttt{AttributeError: module `os' has no attribute `geteuid'}), 
unavailable modules (\eg{}, \texttt{ModuleNotFoundError: No module named `fcntl'}), 
encoding errors (\texttt{UnicodeDecodeError: `charmap' codec...}), 
specialized exceptions like \texttt{TclError} for display issues or 
\texttt{ZoneInfoNotFoundError} for timezone problems, and platform-specific errors 
like \texttt{OSError: [Errno 24] Too many open files} for file descriptor limits.
These signatures facilitate more direct diagnosis from logs, reducing the need for 
extensive code inspection required by context-dependent symptoms.

\MyPara{Context-dependent symptoms}~The remaining 9 sub-categories (37.5\%) 
do not produce unique error signatures, requiring developers to inspect the code 
context to diagnose the portability issue. These include cases where the same error 
can arise from multiple root causes, where failures are silent or produce generic errors, 
or where the symptom manifests as behavioral differences rather than exceptions. 
For example, path separator issues often manifest as generic \texttt{FileNotFoundError} 
or silent failures where hardcoded paths like \texttt{"/tmp/file.txt"} work on Unix but 
fail on Windows. Similarly, case sensitivity issues may produce \texttt{FileNotFoundError} 
indistinguishable from actual missing files unless one examines the filesystem. 
The /proc filesystem is another case where generic \texttt{FileNotFoundError} occurs on 
macOS/Windows, requiring code inspection to determine if the issue stems from missing /proc 
paths rather than other file access problems. Shell command execution issues rarely throw 
exceptions but instead manifest as incorrect exit codes or output differences when commands 
are interpreted by different shells (\texttt{cmd.exe} vs. \texttt{bash}). GUI differences 
show as visual rendering variations or behavioral inconsistencies rather than explicit errors, 
and terminal capability mismatches may cause ANSI color codes to appear as literal text. 
Permission issues vary depending on the specific operation (chmod, file access, etc.), 
filesystem block size problems appear as test assertions on exact byte counts that differ 
across platforms, and binary wheel mismatches often succeed during installation but fail at 
runtime with ABI compatibility errors that vary by architecture.

\MyPara{Implications for diagnosis}~Our findings reveal that while
unique signatures enable rapid automated triage for 62.5\% of
portability issues, the remaining 37.5\% require more sophisticated
analysis. This suggests that detection tools must go beyond simple log
parsing and incorporate contextual hints (\eg{}, hardcoded paths,
shell command patterns, or platform-specific API usage). The presence
of context-dependent symptoms also highlights the importance of
comprehensive cross-platform testing, as these issues cannot be
reliably predicted through static analysis alone.

\begin{tcolorbox}[boxrule=0pt,sharp corners,boxsep=1pt,left=2pt,right=2pt,top=2pt,bottom=1pt]
\textbf{RQ\textsubscript{2.2} (symptoms)}: Of 24 sub-categories, 15 (62.5\%) produce unique error signatures enabling
 immediate diagnosis, while 9 (37.5\%) exhibit context-dependent symptoms requiring code inspection to distinguish
  from other failure modes.
\end{tcolorbox}
 
\subsubsection{\textbf{Answering \RQTwoThree{}}}
\label{rq2-3}

\begin{figure}[]
\centering
\begin{subfigure}{0.48\textwidth}
\begin{lstlisting}[caption={Defensive check for optional library}, label={lst:defensive}]
- import readline
+ try:
+     import readline
+ except ImportError:
+     readline = None  # fallback if readline is unavailable
\end{lstlisting}
\end{subfigure}
\begin{subfigure}{0.48\textwidth}
\begin{lstlisting}[caption={Environment handling: Using pytest skip}, label={lst:environment}]
+ import platform
+ @pytest.mark.skipif(platform.system() == "Darwin", reason="Not supported on macOS")
def test_special_case():
    ...
\end{lstlisting}
\end{subfigure}

\begin{subfigure}{0.48\textwidth}
\begin{lstlisting}[caption={Normalization: Normalizing file paths}, label={lst:normalization}]
- file_path = "data/file.txt"
+ import os
+ file_path = "data" + os.path.sep + "file.txt"  # consistent path
\end{lstlisting}
\end{subfigure}

\begin{subfigure}{0.48\textwidth}
\begin{lstlisting}[caption={Portable APIs: Using platform module for OS detection}, label={lst:portable}]
- hostname = os.uname()[1]  # unix-only function
+ import platform
+ hostname = platform.uname().node  # cross-platform
\end{lstlisting}
\end{subfigure}
\caption{Representative examples of fixes for portability issues: (a) Defensive checks, (b) Environment handling, (c) Normalization, and (d) Portable APIs.}
\label{fig:patterns}
\end{figure}

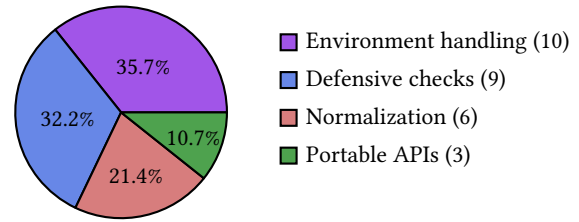
\begin{figure}[t!]
\centering
\begin{tikzpicture}
\pie[
    text=legend,
    radius=1.4,
    color={EnvColor!80,DefensiveColor!80,NormColor!80,PortableColor!80}
]{
    35.7/Environment handling (10),
    32.2/Defensive checks (9),
    21.4/Normalization (6),
    10.7/Portable APIs (3)
}
\end{tikzpicture}
\caption{\label{fig:fix_pattern_distribution}Distribution of fix patterns across 24 portability
  sub-categories.}
\vspace{-6ex}
\end{figure}

\noindent

\MyPara{Fix patterns}~By examining source code, test logs, issue discussions, and 
merged pull requests from both detection approaches, we identified four general fix 
patterns that address portability problems across all 24 sub-categories (mapped in 
Table~\ref{tab:symptoms}). Figure~\ref{fig:patterns} illustrates these patterns with 
concrete examples:

\begin{packed_enumerate}
\item \textbf{Environment handling} adapts code and tests to runtime conditions by 
detecting platform capabilities and adjusting behavior. Listing~\ref{lst:environment} 
shows this pattern using pytest's \texttt{skipif} decorator to conditionally skip 
tests on macOS. This allows test suites to adapt to platform-specific capabilities 
without maintaining separate test files for each operating system. This is the most 
prevalent pattern, accounting for 35.7\% of all pattern applications.

\item \textbf{Defensive checks} add guards, fallbacks, or conditional imports to handle 
missing APIs or modules gracefully. Listing~\ref{lst:defensive} demonstrates this 
pattern by wrapping the \texttt{readline} import (unavailable on Windows) in a 
\texttt{try/except} block with a fallback value, allowing the code to continue 
execution rather than crash. This pattern applies when functionality may be unavailable 
on certain platforms and accounts for 32.2\% of pattern applications.

\item \textbf{Normalization} standardizes encodings, line endings, paths, or filenames 
to prevent spurious differences across platforms. Listing~\ref{lst:normalization} 
illustrates this by constructing file paths using \texttt{os.path.sep} instead of 
hardcoded separators, ensuring correct behavior on both Unix (forward slash) and 
Windows (backslash) systems. This pattern accounts for 21.4\% of applications.

\item \textbf{Portable APIs} replace OS-specific constructs with cross-platform 
abstractions. Listing~\ref{lst:portable} demonstrates replacing the Unix-only 
\texttt{os.uname()} function with the portable \texttt{platform.uname()} function, 
ensuring the code works correctly on both Unix-like systems and Windows. While 
used less frequently (10.7\%), this pattern targets specific, well-defined problems 
where standard cross-platform abstractions exist.

\end{packed_enumerate}

\MyPara{Pattern
  distribution}~Figure~\ref{fig:fix_pattern_distribution} shows how
these patterns distribute across all pattern applications. Some
sub-categories require multiple patterns, resulting in 28 total
pattern applications across 24 sub-categories. Environment handling is
the most prevalent strategy, accounting for 10 applications (35.7\%),
followed by Defensive checks with 9 applications (32.2\%), together
representing approximately two-thirds of all pattern applications.
Normalization accounts for 6 applications (21.4\%), primarily
addressing file and path-related issues. Portable APIs, while used
least frequently (3 applications, 10.7\%), targets specific,
well-defined problems where standard cross-platform abstractions
exist. This distribution reflects the nature of portability issues:
most problems arise from platform-environmental differences
(explaining fixes with conditional handling) rather than fundamental
API incompatibilities. The data also show that portability fixes are
typically short and localized rather than extensive or
widespread. Consequently, they tend to impose a low workload on
developers for both implementation and review.

\begin{tcolorbox}[boxrule=0pt,sharp corners,boxsep=2pt,left=2pt,right=2pt,top=2.5pt,bottom=2pt]
\textbf{RQ\textsubscript{2.3} (fixes)}: We identified \NumFixPatterns{} general fix 
patterns that collectively address all \NumRootCauseSubcategories{} portability 
sub-categories. Some sub-categories require multiple patterns, resulting in 28 total 
pattern applications. Environment handling is most prevalent (35.7\%), followed by 
Defensive checks (32.2\%), Normalization (21.4\%), and Portable APIs (10.7\%). 
These patterns represent recurring, localized code modifications that provide 
systematic guidance for resolving portability issues across diverse contexts.
\end{tcolorbox}

\subsection{\textbf{Answering} \RQThree{}}
\label{rq3}

This section evaluates performance of tools for detecting
(\textsection~\ref{sec:llm-detection}) and for fixing
(\textsection~\ref{sec:llm-fixing}) portability issues.

\begin{table}[t!]
\centering
  \caption{~\label{tab:llm_detection_results} LLMs for detection: metrics.}
  \vspace{-2ex}  
  \setlength{\tabcolsep}{1.5pt} %
  \begin{subtable}[t]{0.5\textwidth}
    \centering
\caption{\label{tab:confusion_matrices_combined}Confusion matrices w/ 90 samples: 60 portable and 30 non-portable.}
\begin{tabular}{lccc|ccc|ccc}
  \toprule
  & \multicolumn{9}{c}{\textbf{Predicted}}\\
\cline{2-10}
& \multicolumn{3}{c|}{\texttt{llama-3.3}} &
\multicolumn{3}{c|}{\texttt{grok-4-fast}} &
\multicolumn{3}{c}{\texttt{gpt-4o-mini}} \\
\textbf{Actual} & \textbf{Port} & \textbf{NonPort} & & \textbf{Port} & \textbf{NonPort} & & \textbf{Port} & \textbf{NonPort} & \\
\midrule
\textbf{Port}     & 10 & 50 & & 53 & 7  & & 21 & 39 & \\
\textbf{NonPort}  & 4  & 26 & & 12 & 18 & & 1  & 29 & \\
\bottomrule
\end{tabular}
  \end{subtable}
  \vspace{1ex}
  \hfill
  \begin{subtable}[t]{0.5\textwidth}
    \centering
\caption{Performance metrics.}
\label{tab:llm_performance}
\begin{tabular}{lrrrrr}
\toprule
Model & Precision & Recall & F1-Score & Accuracy \\
\midrule
\texttt{llama-3.3} & 0.34 & 0.87 & 0.49 & 0.40 \\
\texttt{grok-4-fast} & 0.72 & 0.60 & 0.65 & 0.79 \\
\texttt{gpt-4o-mini} & 0.43 & 0.97 & 0.59 & 0.56 \\
\bottomrule
\end{tabular}
  \end{subtable}
\end{table}

\begin{table}[t!]
\centering
\caption{Prompts used for LLM evaluation.}
\label{tab:llm_prompts}

\begin{subtable}[t]{0.97\columnwidth}
\centering
\caption{Generic prompt used for portability detection and fixing.}
\label{tab:generic_prompt}
\begin{tabular}{p{0.97\columnwidth}}
\toprule
\textbf{Generic Prompt} \\
\midrule
You are a Python expert. Check the following code and answer: \\
1. Is there any operation in the code that could fail on a specific operating system (Linux, Mac, Windows)? \\
2. If yes, explain why and on which OS it might fail. If it is fully portable, finish saying ``Portable''. \\
\bottomrule
\end{tabular}
\end{subtable}

\vspace{1ex}

\begin{subtable}[t]{0.97\columnwidth}
\centering
\caption{Pattern-guided prompt template to fix portability issues.}
\label{tab:specific_prompts}
\begin{tabular}{p{0.97\columnwidth}}
\toprule
\textbf{Pattern-Guided Prompt Template} \\
\midrule
You are a Python expert. The following code has a portability problem:
\textit{<symptom-description>}. Fix the code, considering the
following fix options \textit{<list-of-possible-fixes-for-the-symptom>}. \\
\bottomrule
\end{tabular}
\end{subtable}
\end{table}

\subsubsection{LLMs for detection}\sloppy
\label{sec:llm-detection}
\textbf{Alternative static analysis tools.} We examine the
capabilities of various popular linters for Python, namely
Ruff~\cite{ruffref}, Flake8~\cite{flake8}, Pylint~\cite{pylint},
MyPy~\cite{mypy}, Bandit~\cite{bandit}, and Pyright~\cite{pyright}.
We find that only Ruff provides a rule that directly addresses
potential portability concerns, namely the
\texttt{unspecified-encoding} check~\cite{encruff} that maps to the
``Encoding'' sub-category within the ``File and Directories''
category. The remaining tools focus primarily on style, type
correctness, security, and general code quality. For that reason, we
explored the use of Large Language Models (LLMs) to detect portability
problems and repair them. We consider three LLMs of moderate model size, namely
\texttt{llama-3.3}~\cite{llama3}, \texttt{Grok-4-Fast}~\cite{grok4},
and \texttt{GPT-4o-Mini}~\cite{gpt4o}

\noindent\textbf{Setup.}~We sampled a total of \NumCodeUsedLLM{} code
snippets to evaluate performance of LLM: (i)~\NumCodeNonPortableByLLM{} non-portable snippets
containing OS-specific constructs, (ii)~\NumCodePortableIssue{} portable snippets that had
been fixed by developers (mined from issues), and (iii)~\NumCodePortableManual{} code we
manually categorized as portable. This distribution attempts to
reflect a more realistic division of code seen in the wild, where
portability violations are less frequent. We used the openrouter.ai
API~\cite{openrouter} to interact with the models. Each model was
prompted with a consistent template that included the code snippet and
a question asking whether the code was portable across Windows, Linux,
and Mac (see Table~\ref{tab:llm_prompts}). We parse the answers to extract 
classification decisions.

\noindent\textbf{Metrics.}~We used standard metrics to measure the performance of binary classifiers (e.g., precision, recall, accuracy, etc.)~\cite{saito2015precision}. Table~\ref{tab:llm_detection_results} shows
results. Table~\ref{tab:confusion_matrices_combined} shows the
confusion matrices associated with each LLM, illustrating the
classification of the \NumCodeUsedLLM{} samples.
Table~\ref{tab:llm_performance} shows the metrics for each model.

\noindent\textbf{Analysis of results.}~ The results we obtain indicate
limited ability of LLMs to accurately reason about portability
issues. llama-3.3 and GPT-4o-Mini are ``pessimistic'' and report too
many alarms, most of which are false positives. In contrast,
Grok-4-Fast is more cautious in reporting but misses lots of true
positives.

\subsubsection{LLMs for fixing.}
\label{sec:llm-fixing}
\noindent\textbf{Setup.}~ Following the classification experiment, we designed a 
complementary evaluation focused on the corrective capabilities of LLMs. In this 
phase, we used the \NumCodeNonPortableByLLM{} non-portable Python snippets identified 
earlier and prompted each model with two distinct repair configurations.  
In the first configuration (\textit{generic prompt}), each model received a general 
instruction stating that the code exhibited portability issues and was asked to 
produce a corrected version (see Table~\ref{tab:llm_prompts}).  
In the second configuration (\textit{pattern-guided prompt}), each non-portable 
snippet was accompanied by the specific symptom (derived from our Unique
signature symptoms in \ref{rq2-2}) and the corresponding \textit{General Fix Pattern} from 
Table~\ref{tab:symptoms}. 

The rationale for this approach is: (1) observe symptoms in the code,
(2) look up the mapping in Table~\ref{tab:symptoms} (Symptom
$\rightarrow$ General Fix Pattern), and (3) prompt the LLM with
specific fix pattern guidance. For example, when encountering code
that produces \texttt{AttributeError: module `signal' has no attribute
  `SIGHUP'}, we would provide that symptom together with corresponding
fix(es) (Table~\ref{tab:symptoms}). Table~\ref{tab:llm_prompts} shows
the template structure used for these pattern-guided prompts.  This
design evaluates whether structured repair guidance improves the
model's ability to generate functionally portable code. Each model
attempted to fix all \NumCodeNonPortableByLLM{} snippets, resulting in
90 repair attempts per configuration. We also examine repair performance 
separately for cases with explicit symptom descriptions versus cases with 
empty symptom fields to understand how input characteristics influence model effectiveness.

\noindent\textbf{Metrics.} ~We evaluate the correctness of a generated
fix with two criteria: (i) the modified code must execute successfully
without errors across all target platforms (Linux, macOS, Windows),
and (ii) the fix must preserve the intended functionality of the
original program.  Fixes that met both criteria were considered
\textit{correct}; otherwise, they were labeled as \textit{incorrect}.
Accuracy is the ratio of correct fixes to the total number of repair
attempts.

\begin{table}[t!]
  \centering
\caption{Performance of LLMs in generating portability fixes across both prompt configurations (\NumCodeNonPortableByLLM{} samples per model).}
\label{tab:llm_fix_results_combined}
\begin{subtable}[t]{0.47\textwidth}
  \centering
\caption{Generic prompt.}
\begin{tabular}{lrrr}
\toprule
\textbf{Model} & \textbf{Correct} & \textbf{Incorrect} & \textbf{Accuracy} \\
\midrule
\texttt{grok-4-fast} & 13 & 17 & 0.43 \\
\texttt{gpt-4o-mini} & 10 & 20 & 0.33 \\
\texttt{llama-3.3} & 8 & 22 & 0.27 \\
\bottomrule
\end{tabular}
\end{subtable}
\hfill
\begin{subtable}[t]{0.47\textwidth}
\centering
\caption{Pattern-guided prompt.}
\begin{tabular}{lrrr}
\toprule
\textbf{Model} & \textbf{Correct} & \textbf{Incorrect} & \textbf{Accuracy} \\
\midrule
\texttt{grok-4-fast} & 23 & 7 & 0.77 \\
\texttt{gpt-4o-mini} & 21 & 9 & 0.70 \\
\texttt{llama-3.3} & 15 & 15 & 0.50 \\
\bottomrule
\end{tabular}
\end{subtable}
  \vspace{-1ex}
\end{table}

\noindent\textbf{Analysis of results.}~ The results in Table~\ref{tab:llm_fix_results_combined} show that all models exhibited limited success under the generic prompt setting, with accuracies ranging from 27\% to 43\%.  
This indicates that, when provided with only minimal context, most LLMs struggle to infer the exact nature of portability issues and to produce appropriate cross-platform corrections.  
While the \texttt{grok-4-fast} model outperformed the others, its success rate remained below 50\%, suggesting that general-purpose reasoning alone is insufficient for precise code repair in this domain.

In contrast, the pattern-guided configuration substantially improved repair performance across all models.  
The inclusion of explicit problem descriptions and generalized fix patterns increased accuracy by more than 30 percentage points on average.  
The \texttt{grok-4-fast} model achieved the highest accuracy (76.67\%), demonstrating a clear ability to integrate structured repair hints into its reasoning process.  
Similarly, \texttt{gpt-4o-mini} reached 70\%, producing fixes that were more consistent with the intended patterns while preserving functional correctness.  
The \texttt{llama-3.3} model also showed moderate improvement (50\%), particularly in cases involving straightforward path and API adjustments, though it still struggled with more complex logic integration.

Table~\ref{tab:llm_fix_results_symptom_breakdown} reports on an
experiment where we analyzed repair performance separately for cases with explicit 
symptom descriptions versus cases lacking such descriptions due to context-dependent symptoms. 
Results indicate that the availability of symptom descriptions significantly influences LLM performance.  
Among the 23 cases with explicit symptom descriptions, \texttt{grok-4-fast} achieves notably higher accuracy (0.83), while \texttt{gpt-4o-mini} and \texttt{llama-3.3} maintain similar or slightly improved performance (0.70 and 0.52, respectively).  
In contrast, for the 7 cases where symptom descriptions were not available due to context-dependent or unstable symptoms, the prompt contained only the code snippet and general instructions to apply the relevant fix pattern. In these cases, performance generally declines: \texttt{grok-4-fast} drops to 0.57, while \texttt{gpt-4o-mini} remains robust at 0.71 and \texttt{llama-3.3} falls to 0.43.  
This pattern demonstrates that explicit symptom descriptions provide valuable context that particularly benefits more sophisticated models, whereas some cases can still be addressed even without rich contextual information.

\begin{table}[t!]
\centering
\caption{Pattern-guided repair performance breakdown: cases with symptom descriptions vs.\ empty symptom fields.}
\label{tab:llm_fix_results_symptom_breakdown}
\begin{subtable}[t]{0.47\textwidth}
\centering
\caption{With symptom descriptions ($n=23$).}
\begin{tabular}{lrrr}
\toprule
\textbf{Model} & \textbf{Correct} & \textbf{Incorrect} & \textbf{Accuracy} \\
\midrule
\texttt{grok-4-fast} & 19 & 4 & 0.83 \\
\texttt{gpt-4o-mini} & 16 & 7 & 0.70 \\
\texttt{llama-3.3} & 12 & 11 & 0.52 \\
\bottomrule
\end{tabular}
\end{subtable}
\hfill
\begin{subtable}[t]{0.47\textwidth}
\centering
\caption{Without symptom descriptions ($n=7$).}
\begin{tabular}{lrrr}
\toprule
\textbf{Model} & \textbf{Correct} & \textbf{Incorrect} & \textbf{Accuracy} \\
\midrule
\texttt{grok-4-fast} & 4 & 3 & 0.57 \\
\texttt{gpt-4o-mini} & 5 & 2 & 0.71 \\
\texttt{llama-3.3} & 3 & 4 & 0.43 \\
\bottomrule
\end{tabular}
\end{subtable}
  \vspace{-3ex}
\end{table}

Overall, these findings highlight the strong impact of guided contextualization on LLM-based repair.  
When supplied with explicit repair patterns, the models produced significantly more reliable and semantically consistent fixes, confirming that structured guidance can substantially enhance cross-platform reasoning capabilities in code generation tasks.

\begin{tcolorbox}[boxrule=0pt,sharp corners,boxsep=1pt,left=1pt,right=1pt,top=1pt,bottom=1pt]
  \textbf{RQ\textsubscript{3} (tool effectiveness):}~ Static analysis tools 
  show poor ability to detect portability problems. LLMs achieve moderate detection 
  accuracy (40--79\% across three models), with substantial variation in precision 
  and recall. For repair tasks, LLMs show limited success with generic prompts 
  (27--43\% accuracy) but improve significantly with pattern-guided prompts 
  (50--77\% accuracy), yielding an average improvement of 30 percentage points. 
  These results show that structured guidance using error signatures and 
  fix patterns (Table~\ref{tab:symptoms}) substantially enhances LLM-based 
  repair.
\end{tcolorbox}

\subsection{\textbf{Answering} \RQFour{}}

This section discusses the pull requests (PRs) we submitted.
\label{sec:pr-discussion}

\subsubsection{Method}
We use the cross-OS test re-execution method
(\textsection~\ref{sec:methods}) to detect portability problems in our
\emph{application set} (\textsection~\ref{sec:projects}). For each
identified issue, we manually apply one of the four fix patterns from our
taxonomy~(Table~\ref{tab:symptoms}). We then validate each fix by re-running 
the affected tests on all three operating systems (Linux, macOS, Windows) to 
confirm consistent behavior across platforms. Once the fix is validated, we 
submit a pull request to the project, providing a clear description of the 
portability issue, the root cause, and an explanation of how the proposed change 
resolves the problem.

\subsubsection{Results}

We detected 47 portability issues; \TotalPROpened{} PRs were submitted
(70\%), with 14 not submitted due to project policies. All of the PRs
match one of the four fix patterns we discussed in Section~\ref{rq2-3}. Table~\ref{tab:prs}
describes the PRs. Of the \TotalPROpened{} pull requests submitted, \TotalPRAccepted{} were accepted at the time of writing. The overall acceptance rate of \TotalPRAprovedPercent{},
along with a zero rejection rate, indicates that the issues we
identified correspond to genuine portability problems recognized by
maintainers. Notably, ``File and Directories'' issues were the
most frequent (15 PRs), reflecting the prevalence of path, encoding,
and file handling inconsistencies across platforms. 
``Process and Signals'' issues exhibited the highest acceptance rate~(100\%), likely due to their critical impact on application
functionality.
Considering the location of the fix, we observe that in
\TotalPRInTestsCode{} cases only the test code changes, in
\TotalPRInProgramCode{} cases only the program code changes, and in
\TotalPRInBothCode{} cases both the test and program code
changes. Note that many cases require changes to the application logic
to ensure consistent behavior across platforms.

Next, we highlight three representative pull requests that illustrate
how portability issues manifest in practice and how our identified fix
patterns address them. These examples span multiple categories and demonstrate
different sub-categories and fix patterns from our taxonomy.

\begin{table}[h]
    \centering
    \small
    \setlength{\tabcolsep}{2pt}
    \caption{
      Summary of Pull Requests (PRs) by issue type, showing opened, accepted, and rejected. The last row shows totals.}
    \label{tab:prs}
    \begin{tabular}{lccc}
        \toprule
        \textbf{Category} & \textbf{Opened} & \textbf{Accepted} & \textbf{Rejected} \\
        \midrule
        File and Directories & 15 & 9 & 0 \\
        API Availability & 12 & 4 & 0 \\
        Process and Signals & 2 & 2 & 0 \\
        System Information & 2 & 1 & 0 \\
        Library Dependencies & 1 & 1 & 0 \\
        Environment and Display & 1 & 0 & 0 \\
        \midrule
        \textbf{Total} & \textbf{\TotalPROpened} & \textbf{\TotalPRAccepted} & \textbf{\TotalPRRejected} \\
        \bottomrule
    \end{tabular}
\end{table}
\vspace{-1ex}

\subsubsection{The webassets Pull Request}

We illustrate how portability issues manifest and are fixed using the \textit{webassets} project~\cite{webassets}, a Python library for managing and bundling web assets such as CSS and JavaScript, with 932 stars.
The problem we encountered falls into the \textbf{File and Directories
  (FILE)} category, specifically the ``File locking''
sub-category. The issue occurred in the test suite where
\CodeIn{tempfile.NamedTemporaryFile} was used without disabling
automatic deletion, causing failures on Windows due to
platform-specific file locking
behavior~\cite{pythonnamedtemporaryfile}. On Windows, the operating
system locks open files more aggressively than on Unix-like systems,
preventing the same file from being reopened while still in use.
Listing~\ref{lst:webassets_fix} shows the diff of our fix applied to
the test file. The original code created a temporary file within a
context manager that automatically deleted the file upon exiting the
\texttt{with} block. However, on Windows, attempting to reopen this
file (which the test needed to do because it's a fixture) failed
because the file was still locked by the previous file handle.

\begin{lstlisting}[caption={Approved PR for temporary file handling issue.}, label={lst:webassets_fix}]
 @pytest.fixture
 def tmp_file():
-  with tempfile.NamedTemporaryFile(mode="wt") as f:
+  with tempfile.NamedTemporaryFile(mode="wt", delete=False) as f:
     for _ in range(100):
       f.write("\n")
     f.flush()
-    yield f.name
+    tmp_path = f.name  # store path before file is closed
+  yield tmp_path
+  os.remove(tmp_path)  # clean up after the test
\end{lstlisting}

Our fix addresses this issue by modifying line 4 to disable automatic
deletion using \CodeIn{delete=False}, storing the file path on line~9
before the file handle is closed, and manually cleaning up the
temporary file on line 11 after the test completes. This pattern
exemplifies the \textbf{Environment handling} fix strategy, where platform-specific behavior is handled through
careful resource management.
This case demonstrates how seemingly simple operations like temporary
file creation can have subtle portability implications that only
manifest under specific platform conditions, highlighting the
importance of cross-platform testing for robust software development.

\subsubsection{The hosteurope-letsencrypt Pull Request}

We demonstrate another portability fix through the hosteurope-letsencrypt project~\cite{hosteuropeletsencrypt}, which has 67 stars and involves system administration functionality. 
The problem falls into the \textbf{API Availability (API)} category. 
The issue is due to the use of \CodeIn{os.geteuid()} to check for 
root privileges, but this function is unavailable on Windows, causing 
\texttt{AttributeError: module `os' has no attribute `geteuid'}.
\begin{lstlisting}[caption={Cross-platform admin privilege detection fix.}, label={lst:shared_fix}]
+import platform
+def is_admin():
+    try:
+        if platform.system() == "Windows":
+            import ctypes
+            return ctypes.windll.shell32.IsUserAnAdmin() != 0
+        else:
+            return os.geteuid() == 0
+    except:
+        return False

-is_root = os.geteuid() == 0
+is_root = is_admin()
\end{lstlisting}

Our fix creates a cross-platform \CodeIn{is\_admin()} function that 
replaces the Unix-only \CodeIn{os.geteuid()} call. The fix uses 
\textbf{Portable APIs} (line 4: \CodeIn{platform.system()} for OS detection), 
\textbf{Environment handling} (lines 6, 8: conditional logic that calls 
\CodeIn{ctypes.windll.shell32.IsUserAnAdmin()} for Windows or 
\CodeIn{os.geteuid()} for Unix), and \textbf{Defensive checks} 
(lines 3--9: try-except wrapper with safe fallback).

\subsubsection{The pydantic-extra-types Pull Request}

We show an example from the pydantic-extra-types
project~\cite{pydantic_extra_types}, a library that provides
additional type validators for Pydantic with 290 stars, used for
extending Pydantic's~\cite{pydantic_validation} validation
capabilities with specialized data types.
The problem we encountered also falls into the \textbf{File and
  Directories (FILE)} category, specifically the ``Path separators''
sub-category. The issue occurred in test fixtures where
\CodeIn{os.path.relpath} was used to create relative paths from
absolute paths. On Windows, this function fails when the source and
target paths are on different drives (e.g., \texttt{C:\textbackslash}
vs \texttt{D:\textbackslash}), as Windows cannot create relative paths
between different drive letters~\cite{python-relpath}.
Listing~\ref{lst:pydantic_fix} shows our fix applied to test
fixtures. The original code assumed that \CodeIn{os.path.relpath}
would always succeed, but on Windows this assumption may not hold. For
example, when tests are executed from a project directory located on
drive \texttt{D:} while the system temporary directory (used to create
fixture files) resides on drive \texttt{C:}, the function call fails
because it attempts to compute a relative path across drives. Since
the location of temporary directories on Windows is determined by
environment variables and may differ from the working
drive~\cite{python-tempfile}, this behavior can lead to path
resolution errors.

\begin{lstlisting}[caption={\vspace{-0.5ex}Patch for cross-drive relative path issue.}, label={lst:pydantic_fix}]
 @pytest.fixture
 def relative_file_path(absolute_file_path: Path) -> Path:
+ cwd = Path.cwd()
+ if os.name == 'nt' and absolute_file_path.anchor != cwd.anchor:
+   return absolute_file_path
  return Path(os.path.relpath(absolute_file_path, os.getcwd()))
\end{lstlisting}

Our fix addresses this Windows-specific limitation by detecting when the OS is Windows (line 4) and checking whether the absolute path and current working directory have different anchors (drive letters). When this condition is met, the function returns the absolute path instead of attempting the problematic relative path conversion (line 5). This pattern exemplifies the \textbf{Environment handling} fix strategy from our taxonomy, where platform-specific limitations are accommodated through conditional logic.

This case demonstrates how path operations that work seamlessly on Unix-like systems can encounter fundamental limitations on Windows due to the multi-drive architecture, highlighting the importance of considering platform-specific filesystem constraints in cross-platform development.

\begin{tcolorbox}[boxrule=0pt,sharp corners,boxsep=2pt,left=2pt,right=2pt,top=2.5pt,bottom=2pt]
   \textbf{RQ\textsubscript{4} (community validation):}~ We submitted
   \TotalPROpened{} pull requests (PRs) to open-source projects,
   achieving a \TotalPRAprovedPercent{} acceptance rate with no rejections. Issues related
   to files and directories were the most common (15 PRs), whereas
   process and signal issues had the highest acceptance rate
   (100\%). These results confirm our initial observation that fixes
   generally involve localized changes—either to test code (\TotalPRInTestsPercent{}) or
   application logic (\TotalPRInProgramPercent{})—and suggest a positive response from
   developers to the submitted PRs.
\end{tcolorbox}

\section{Lessons Learned}
\label{sec:lessons}

This section distills key insights from our study into actionable guidance 
for developers working on cross-platform Python projects.

\begin{packed_enumerate}
  \item \textbf{Porting code remains important despite the rise of
  containerization technology.} We find that a relatively small
fraction of open-source projects (19.9\%) from our dataset include
Docker configurations. The majority of projects need to handle
portability natively in code.
\emph{Recommendations.}~(1)~Maintain cross-OS CI test
matrices~\cite{github2025matrix} to catch issues early, regardless of
deployment target; (2) Fine tune code
agents~\cite{microsoft2025copilot} to detect (and fix) portability
issues as code is written.

  \item \textbf{Guided LLM repair using error signatures and fix patterns.} 
LLMs underperform with generic ``fix this code'' prompts (27--43\% success) but 
achieve 50--77\% success when guided with specific error messages and corresponding 
fix patterns from Table~\ref{tab:symptoms}.
\emph{Recommendations.}~Include the specific error message and suggest the 
corresponding fix pattern in your LLM prompt. This structured approach nearly 
doubles success rates compared to generic prompts.

  \item \textbf{Portability fixes are localized, not architectural.} 
Our four fix patterns (Table~\ref{tab:symptoms}) address issues through small, 
targeted modifications—adding conditional imports, normalizing paths, or wrapping 
platform-specific calls—rather than redesigning system architecture.
\emph{Recommendations.}~Treat portability issues as localized problems. Consult 
Table~\ref{tab:symptoms} to identify the issue type and apply the corresponding 
fix pattern rather than deferring fixes as ``too invasive''.

  \item \textbf{File and path issues dominate cross-platform failures.} 
File and Directories is the most prevalent category, affecting 62 projects (41\% 
of all portability issues). Path separators, encoding, and file locking are the 
most common sub-categories.
\emph{Recommendations.}~Use \texttt{pathlib.Path} for path construction, explicitly 
specify \texttt{encoding} when opening text files, and use context managers 
(\texttt{with} statements) to avoid Windows file locking issues.
\end{packed_enumerate}

\section{Threats to Validity}
\label{sec:threats}

\noindent\textbf{Internal validity.} Our cross-OS test re-execution relies on GitHub
Actions runners, which may differ from physical machines. In addition,
the manual analysis of GitHub issues involved subjective
interpretation. Mitigation: multiple researchers independently
conducted the manual analysis, and disagreements were resolved through
consensus.

\vspace{0.5ex}
\noindent\textbf{External validity.} The 2,042 GitHub projects we
analyzed may not fully represent all software domains or
domain-specific applications. Mitigation: we sampled projects of
diverse domains and sizes. Our taxonomy is derived from fundamental
operating-system differences and thus transcends specific
domains. Furthermore, the high pull-request acceptance rate supports
the generalizability of our findings across multiple developer
communities.

\vspace{0.5ex}
\noindent\textbf{Conclusion validity.} Sample sizes varied across
research questions (500, 400, and 90 snippets), which may affect the
reliability of cross-question comparisons. Mitigation: we triangulated
results using multiple independent methods, validated the taxonomy
with both detection approaches, and obtained a \TotalPRAprovedPercent{} pull-request
acceptance rate with zero rejections.

Finally, our scope is limited to Linux, macOS, and Windows, which are
representative platforms. We did not consider portability across
hardware architectures or Python versions, for instance.

\section{Related Work}
\label{sec:related}

Our work relates to prior research on software portability and cross-platform testing, but differs in scope and methodology.

\subsection{Empirical Studies on Software Portability}~Ghandorh \etal{}~\cite{ghandorh2020systematic} performed a systematic literature review of metrics and methodologies for assessing software portability. 
Their work surveys existing approaches for measuring portability but does not provide empirical data on real-world portability issues or concrete solutions. 
In contrast, our study provides the first large-scale empirical characterization of Python portability problems through systematic testing and issue analysis, developing both a comprehensive taxonomy and practical repair patterns. 
While their review identifies the fragmented nature of portability research, our work addresses this gap by establishing a unified framework specifically for Python cross-platform issues.

Wang \etal{}~\cite{wang2023compatibility} present an empirical study of compatibility issues in deep learning systems, analyzing thousands of posts to categorize common problems and their causes. 
While their findings relate to cross-environment inconsistencies, the scope is broader than portability and centered on DL ecosystems. 
Our work instead focuses directly on Python cross-platform portability, providing a targeted taxonomy and concrete repair strategies.

\vspace{-1ex}
\subsection{Cross-Platform Testing and Defect Studies}~Vahabzadeh \etal{}~\cite{vahabzadeh2015empirical} conducted an empirical study on test code defects, finding that approximately 18\% of test bugs stem from environmental factors, particularly Windows vs. Unix platform differences. 
While their work acknowledges environment-related testing issues, it focuses on general test code quality and defect classification rather than systematically characterizing portability problems or developing repair strategies. 
Our work differs in three key ways: (1) we provide the first comprehensive taxonomy of Python portability issues with 7 categories and 24 sub-categories, (2) we develop systematic fix patterns that address these issues, and (3) we evaluate both detection and repair approaches using modern tools (LLMs) that were not available during their study.

\vspace{-1ex}
\subsection{Dynamic Analysis for Portability}~Rasley \etal{}~\cite{rasley2015detecting} proposed \CodeIn{CheckAPI}, a runtime framework that detects cross-platform API violations by comparing execution traces with platform-specific specifications.
Unlike CheckAPI, which requires predefined API models and targets API-level compatibility, our work observes behavioral differences directly through cross-OS re-execution and covers broader portability issues such as file systems, encodings, dependencies, and environments.

\vspace{-1ex}
\subsection{API Specification Nondeterminism}~Shi \etal{}~\cite{ShiETAL16NonDex} and Gyori \etal{}~\cite{gyori2016nondex} introduced \textit{NonDex}, a tool that systematically explores alternative valid behaviors of Java APIs to uncover bugs caused by deterministic assumptions about inherently nondeterministic specifications (e.g., iteration order in hash-based collections).
Although NonDex focuses on intra-platform variability within the JVM, our study examines deterministic divergences that emerge across different operating systems.
Both approaches highlight how subtle, often implicit, assumptions about platform behavior can compromise software reliability under diverse execution environments.

\vspace{-1ex}
\subsection{OS Variability and System Reliability}~Sun \etal{}~\cite{sun2016bear} proposed \textit{Bear}, a framework that quantifies how nondeterministic OS behaviors (e.g., scheduling and I/O timing) affect application reliability. 
They show that even minor OS-level variations can propagate to failures or performance drops, revealing hidden fragility in applications that assume predictable OS behavior. 
While Bear focuses on statistical modeling of OS nondeterminism, our work addresses deterministic portability issues stemming from platform-specific APIs, file systems, and libraries, providing concrete fix patterns through cross-platform testing.

\vspace{-1ex}
\subsection{Static Analysis for Python}~Popular tools such as Ruff~\cite{ruffref}, Flake8~\cite{flake8}, Pylint~\cite{pylint}, MyPy~\cite{mypy}, Bandit~\cite{bandit}, and Pyright~\cite{pyright} focus on style, typing, and security, offering minimal support for portability detection—only Ruff includes \texttt{unspecified-encoding}~\cite{encruff}.
Specialized analyzers like Pysa~\cite{pysa2021} target security (e.g., taint and injection). Our work instead addresses runtime portability issues caused by OS variability, mostly missed by existing tools.

\vspace{-1ex}
\subsection{Performance and Portability.}~
Zioga \etal~\cite{ziogas2021productivity} analyzed performance bugs
across architectures. Our study differs from theirs in two
aspects. First, we focus on portability issues across operaring
systems not architectures. Second, we focus on functional correctness
not performance. Awar \etal{} proposes framework to automatically
translate Python in C++ code~\cite{kokkos}. The work is orthogonal to
ours as we do not focus on efficiency gains.

\section{Conclusions and Future Work}
\label{sec:conclusions}

This paper provides the first comprehensive characterization of
cross-OS portability issues in Python code. We employ two methods to
find these issues: cross-OS test re-execution and issue mining.  We
analyze 900 projects and find that 16.8\% exhibit portability
problems.
Issues related to File and Directory operations are the most
prevalent. Overall, our taxonomy captures seven categories of root
causes of issues (with 24 sub-categories) and four fix patterns for
those issues.  We find that LLMs achieve 40-79\% detection accuracy
and 50-77\% repair success when guided by structured patterns. Our
validation through 33 pull requests showed an acceptance rate of
\TotalPRAprovedPercent{}, indicating that developers care about these
issues.

Future work can extend this study by developing hybrid detection tools
that integrate cross-OS re-execution, symptom signatures, and
lightweight static analyses. Another promising direction is the
fine-tuning of code agents for portability issues using pattern-guided
prompts. Finally, it is important to broaden the evaluation to cover a
wider range of operating system versions, hardware architectures,
Python versions and even other programming languages.

\vspace{-1ex}
\section*{Data Availability}
The artifacts are publicly available at the following link:\\
\vspace{-3ex}
\begin{center}
  \ourrepo
\end{center}  
\vspace{-1.75ex}

\begin{acks}
\label{sec:acks}
\vspace{-1ex}
This work is partially supported by the NSF under Grant Nos. CCF-2349961 and CCF-2319472. 
Denini was supported by CNPq Grant No. 140220/2022-4.
We thank Mohammed Yaseen for his assistance with experimental setup.
\end{acks}

\onecolumn \begin{multicols}{2}
\bibliographystyle{ACM-Reference-Format}
\bibliography{ref}

@string{FSE = "FSE"}

@String{ICSME = "ICSME"}

@string{ICST = "ICST"}

@string{ISSRE = "ISSRE"}

@misc{Pytest,
  longauthor   = {Holger Krekel and pytest-dev},
  title        = {Pytest},
  key          = {Pytest},
  longyear     = {2022},
  howpublished = {{https://docs.pytest.org/en/7.2.x}}
}

@inproceedings{ShiETAL16NonDex,
  title={Detecting assumptions on deterministic implementations of non-deterministic specifications},
  author={Shi, August and Gyori, Alex and Legunsen, Owolabi and Marinov, Darko},
  booktitle={2016 IEEE international conference on software testing, verification and validation (ICST)},
  pages={80--90},
  year={2016},
  organization={IEEE}
}

@article{paszke2019pytorch,
  title={PyTorch: An imperative style, high-performance deep learning library},
  author={Paszke, Adam and Gross, Sam and Massa, Francisco and Lerer, Adam and Bradbury, James and Chanan, Gregory and Killeen, Trevor and Lin, Zeming and Gimelshein, Natalia and Antiga, Luca and others},
  journal={Advances in neural information processing systems},
  volume={32},
  year={2019}
}

@misc{jax,
    author = "{The JAX authors}",
    title = {{JAX library documentation}},
    howpublished = {\url{https://jax.readthedocs.io/en/latest/quickstart.html}},
    note = {Online; accessed in Oct 2024} ,
    year=2024,
}

@inproceedings{vahabzadeh2015empirical,
  title={An empirical study of bugs in test code},
  author={Vahabzadeh, Arash and Fard, Amin Milani and Mesbah, Ali},
  booktitle={2015 IEEE international conference on software maintenance and evolution (ICSME)},
  pages={101--110},
  year={2015},
  organization={IEEE}
}

@inproceedings{rasley2015detecting,
  title={Detecting latent cross-platform api violations},
  author={Rasley, Jeff and Gessiou, Eleni and Ohmann, Tony and Brun, Yuriy and Krishnamurthi, Shriram and Cappos, Justin},
  booktitle={2015 IEEE 26th International Symposium on Software Reliability Engineering (ISSRE)},
  pages={484--495},
  year={2015},
  organization={IEEE}
}

@inproceedings{sun2016bear,
  title={Bear: A framework for understanding application sensitivity to os (mis) behavior},
  author={Sun, Ruimin and Lee, Andrew and Chen, Aokun and Porter, Donald E and Bishop, Matt and Oliveira, Daniela},
  booktitle={2016 IEEE 27th International Symposium on Software Reliability Engineering (ISSRE)},
  pages={388--399},
  year={2016},
  organization={IEEE}
}

@inproceedings{ghandorh2020systematic,
  title={A systematic literature review for software portability measurement: Preliminary results},
  author={Ghandorh, Hamza and Noorwali, Abdulfattah and Nassif, Ali Bou and Capretz, Luiz Fernando and Eagleson, Roy},
  booktitle={Proceedings of the 2020 9th International Conference on Software and Computer Applications},
  pages={152--157},
  year={2020}
}

@misc{stackoverflow2025,
title = {Stack Overflow Developer Survey},
year = {2025},
url = {https://survey.stackoverflow.co/2025/}
}

@misc{githuboctoverse2024,
author = {GitHub Staff},
title = {Octoverse: AI leads Python to top language as the number of global developers surges},
howpublished = {\url{https://github.blog/news-insights/octoverse/octoverse-2024/}},
year = {2024}
}

@misc{pypl2025,
title = {PYPL PopularitY of Programming Language index},
howpublished = {\url{https://pypl.github.io/}},
year = {2025}
}

@article{mince2023,
  title={The Grand Illusion: The Myth of Software Portability and Implications for ML Progress.},
  author={Mince, Fraser and Dinh, Dzung and Kgomo, Jonas and Thompson, Neil and Hooker, Sara},
  journal={Advances in Neural Information Processing Systems},
  volume={36},
  pages={21217--21229},
  year={2023}
}

@misc{ruffref,
author = {Astral Team},
title = {Ruff: A fast Python linter, written in Rust},
year = {2025},
url = {https://docs.astral.sh/ruff/}
}

@misc{flake8,
author = {Ian Stapleton Cordasco and contributors},
title = {Flake8: the modular source code checker for Python},
year = {2025},
url = {https://flake8.pycqa.org/}
}

@misc{pylint,
author = {Logilab and Pylint contributors},
title = {Pylint: Python code static checker},
year = {2025},
url = {https://pylint.pycqa.org/}
}

@misc{mypy,
author = {Jukka Lehtosalo and contributors},
title = {MyPy: Optional static typing for Python},
year = {2025},
url =  {http://mypy-lang.org/}
}

@misc{bandit,
author = {PyCQA},
title = {Bandit: A tool to find common security issues in Python code},
year = {2025},
url = {https://bandit.readthedocs.io/en/latest/}
}

@misc{pyright,
author = {Microsoft Corporation},
title = {Pyright: Static type checker for Python},
year = {2025},
url = {https://github.com/microsoft/pyright}
}

@misc{encruff,
author = {Ruff Team},
title = {Rule Ruff: unspecified encoding},
year = {2025},
url = {https://docs.astral.sh/ruff/rules/unspecified-encoding}
}

@article{kiuwancodeportability,
title = {A Guide to Code Portability},
author = {Kiuwan},
year = {2025},
journal = {Kiuwan Blog},
url = {https://www.kiuwan.com/blog/what-is-code-portability/}
}

@inproceedings{job2024,
 title={Availability and usage of platform-specific APIs: a first empirical study},
  author={Job, Ricardo and Hora, Andre},
  booktitle={Proceedings of the 21st International Conference on Mining Software Repositories},
  pages={27--31},
  year={2024}
}

@misc{github_actions_runners,
  author       = {GitHub},
  title        = {About GitHub-hosted runners},
  year         = {2025},
  url          = {https://docs.github.com/en/actions/using-github-hosted-runners/about-github-hosted-runners},
}

@misc{pytest_tool,
  author       = {Pytest Development Team},
  title        = {pytest: Simple powerful testing with Python},
  year         = {2025},
  url          = {https://docs.pytest.org/en/stable/},
}

@misc{pytest_json_report,
  author       = {Numirias},
  title        = {pytest-json-report: A plugin to generate JSON reports for pytest},
  year         = {2025},
  url          = {https://github.com/numirias/pytest-json-report},
}

@misc{llama3,
  title        = {Llama 3.3 70B Instruct},
  author       = {Meta},
  year         = {2025},
  url          = {https://openrouter.ai/meta-llama/llama-3.3-70b-instruct},
}

@misc{grok4,
  title        = {Grok 4 Fast},
  author       = {xAI},
  year         = {2025},
  url          = {https://openrouter.ai/x-ai/grok-4-fast}
}

@misc{gpt4o,
  title        = {GPT-4o Mini},
  author       = {OpenAI},
  year         = {2025},
  url          = {https://openrouter.ai/openai/gpt-4o-mini}
}

@misc{openrouter,
  title        = {OpenRouter},
  author       = {OpenRouter Team},
  year         = {2025},
  url          = {https://openrouter.ai}
}

@book{strauss1990basics,
  title     = {Basics of Qualitative Research: Grounded Theory Procedures and Techniques},
  author    = {Strauss, Anselm and Corbin, Juliet},
  year      = {1990},
  publisher = {Sage Publications},
  address   = {Newbury Park, CA}
}

@misc{pysa2021,
  author = {Facebook/Meta Engineering},
  title = {Pysa: Security-Focused Static Analysis Tool for Python},
  url = {https://developers.facebook.com/blog/post/2021/04/29/eli5-pysa-security-focused-analysis-tool-python/},
  year = {2021}
}

@misc{webassets,
  title        = {webassets: Asset management for Python web development},
  author       = {Elsdoerfer, Michael},
  year         = {2025},
  howpublished = {\url{https://github.com/miracle2k/webassets}},
}

@misc{hosteuropeletsencrypt,
  title        = {Let's Encrypt Skripte für Hosteurope WebHosting},
  author       = {Sebastian, Stein},
  year         = {2025},
  howpublished = {\url{https://github.com/steinsag/hosteurope-letsencrypt}},
}

@misc{pydantic_extra_types,
  title        = {Pydantic Extra Types},
  author       = {Pydantic Contributors},
  year         = {2025},
  howpublished = {\url{https://github.com/pydantic/pydantic-extra-types}},
}

@misc{pydantic_validation,
  title        = {Pydantic Validation},
  author       = {Pydantic Contributors},
  year         = {2025},
  howpublished = {\url{https://docs.pydantic.dev/latest/}},
}

@misc{python-relpath,
  author       = {Python Software Foundation},
  title        = {os.path.relpath — Return a relative filepath to path},
  year         = {2025},
  howpublished = {\url{https://docs.python.org/3/library/os.path.html\#os.path.relpath}},
}

@misc{python-tempfile,
  author       = {Python Software Foundation},
  title        = {tempfile — Generate temporary files and directories},
  year         = {2025},
  howpublished = {\url{https://docs.python.org/3/library/tempfile.html}},
}

@inproceedings{gyori2016nondex,
  title={NonDex: A tool for detecting and debugging wrong assumptions on Java API specifications},
  author={Gyori, Alex and Lambeth, Ben and Shi, August and Legunsen, Owolabi and Marinov, Darko},
  booktitle={Proceedings of the 2016 24th ACM SIGSOFT International Symposium on Foundations of Software Engineering},
  pages={993--997},
  year={2016}
}

@inproceedings{flaky_tests_empirical,
  title     = {An Empirical Analysis of Flaky Tests},
  author    = {Luo, Qingzhou and Nagappan, Meiyappan and Vasilescu, Bogdan and Malik, H.},
  booktitle = {Proceedings of the 22nd ACM SIGSOFT International Symposium on Foundations of Software Engineering (FSE)},
  year      = {2014}
}

@misc{docker,
  title        = {Docker: Empowering App Development for Developers},
  author       = {{Docker Inc.}},
  howpublished = {\url{https://www.docker.com/}},
  year         = {2013}
}

@inproceedings{apptainer,
  title     = {Singularity: Scientific Containers for Mobility of Compute},
  author    = {Kurtzer, Gregory M. and Sochat, Vanessa and Bauer, Michael W.},
  booktitle = {Proceedings of the 37th International Conference on High Performance Computing, Networking, Storage and Analysis (SC)},
  year      = {2017},
  publisher = {ACM},
  doi       = {10.1145/3126908.3126923},
  note      = {Now maintained as Apptainer: \url{https://apptainer.org/}}
}

@online{microsoft2025copilot,
  author       = {{Microsoft Corporation}},
  title        = {Microsoft 365 Copilot Tuning overview (preview)},
  year         = {2025},
  month        = may,
  day          = {20},
  url          = {https://learn.microsoft.com/en-us/copilot/microsoft-365/copilot-tuning-overview},
}

@online{github2025matrix,
  author       = {GitHub, Inc.},
  title        = {Running variations of jobs in a workflow},
  year         = {2025},
  url          = {https://docs.github.com/actions/writing-workflows/choosing-what-your-workflow-does/running-variations-of-jobs-in-a-workflow},
}

@misc{pythongeteuid,
  title = {os.geteuid() - Return the current process's effective user ID},
  author = {Python Software Foundation},
  year = {2025},
  url  = {https://docs.python.org/3/library/os.html#os.geteuid}
}

@misc{pythonnamedtemporaryfile,
  title = {tempfile.NamedTemporaryFile - Create a named temporary file},
  author = {Python Software Foundation},
  year = {2025},
  url  = {https://docs.python.org/3/library/tempfile.html#tempfile.NamedTemporaryFile}
}

@article{saito2015precision,
  title={The precision-recall plot is more informative than the ROC plot when evaluating binary classifiers on imbalanced datasets},
  author={Saito, Takaya and Rehmsmeier, Marc},
  journal={PloS one},
  volume={10},
  number={3},
  pages={e0118432},
  year={2015},
  publisher={Public Library of Science San Francisco, CA USA}
}

@inproceedings{wang2023compatibility,
  title={Compatibility issues in deep learning systems: Problems and opportunities},
  author={Wang, Jun and Xiao, Guanping and Zhang, Shuai and Lei, Huashan and Liu, Yepang and Sui, Yulei},
  booktitle={Proceedings of the 31st ACM Joint European Software Engineering Conference and Symposium on the Foundations of Software Engineering},
  pages={476--488},
  year={2023}
}

@inproceedings{ziogas2021productivity,
  title={Productivity, portability, performance: Data-centric Python},
  author={Ziogas, Alexandros Nikolaos and Schneider, Timo and Ben-Nun, Tal and Calotoiu, Alexandru and De Matteis, Tiziano and de Fine Licht, Johannes and Lavarini, Luca and Hoefler, Torsten},
  booktitle={Proceedings of the International Conference for High Performance Computing, Networking, Storage and Analysis},
  pages={1--13},
  year={2021}
}

@misc{kokkos,
  title = {Kokkos C++ Performance Portability Programming Ecosystem: The Programming Model - Parallel Execution and Memory Abstraction},
  author = {Kokkos Contributors},
  year = {2025},
  url = {https://github.com/kokkos/kokkos},
}
\end{multicols}

\end{document}